# Inference on differences between classes using cluster-specific contrasts of mixed effects

S.K. Ng [1,*], G.J. McLachlan [2], K. Wang [2], Z. Nagymanyoki [3], S. Liu [3], and S.-W. Ng [3]

[1]School of Medicine, Griffith Health Institute, Griffith University, Meadowbrook, QLD 4131, Australia,
[2]Department of Mathematics, University of Queensland, Brisbane, QLD 4072, Australia,
[3]Laboratory of Gynecologic Oncology, Department of Obstetrics, Gynecology and Reproductive Biology, Brigham and Women's Hospital, Boston, MA 02115, USA

* To whom correspondence should be addressed. Supplementary materials are available on request from the corresponding author.

**Abstract:**

The detection of differentially expressed (DE) genes, that is, genes whose expression levels vary between two or more classes representing different experimental conditions (say, diseases) is one of the most commonly studied problems in bioinformatics. For example, the identification of DE genes between distinct disease phenotypes is an important first step in understanding and developing treatment drugs for the disease. It can also contribute significantly to the construction of a discriminant rule (classifier) for predicting the class of origin of an unclassified tissue sample from a patient. We present a novel approach to the problem of detecting DE genes that is based on a test statistic formed as a weighted (normalized) cluster-specific contrast in the mixed effects of the mixture model used in the first instance to cluster the gene profiles into a manageable number of clusters. The key factor in the formation of our test statistic is the use of gene-specific mixed effects in the cluster-specific contrast. It thus means that the (soft) assignment of a given gene to a cluster is not crucial. This is because in addition to class differences between the (estimated) fixed effects terms for a cluster, gene-specific class differences also contribute to the cluster-specific contributions to the final form of the test statistic. The proposed test statistic can be used where the primary aim is to rank the genes in order of evidence against the null hypothesis of no DE. We also show how a $P$-value can be calculated for each gene for use in multiple hypothesis testing where the intent is to control the false discovery rate (FDR) at some desired level. With the use of real and simulated data sets, we show that the proposed contrast-based approach outperforms other methods commonly used for the detection of DE genes both in a ranking context with lower proportion of false discoveries and in a multiple hypothesis testing context with higher power for a specified level of the FDR.

# 1   Introduction

In the analysis of multivariate feature data from samples in $m$ known classes of $p$ objects, one of the major goals is the inference on differences in observations between classes. This leads to the identification of relevant features that differentiate the classes and the prediction of the class of origin for unclassified objects. An example of such an application is the analysis of gene-expression data, where the aim is to detect genes (features) that are differentially expressed (DE) in a known number $m$ of classes (conditions), $C_1, \ldots, C_m$. These $m$ classes may correspond to tissues (cells) that are at different stages in some process or in distinct pathological states. For example, one can compare healthy cells to cancerous cells to learn which genes tend to be over- or under-expressed in the diseased cells. In this context, the intent is also to select a small subset of "marker" genes that characterize the different tissue classes and construct a classifier to predict the class of origin of an unclassified tissue sample with respect to one of a number of distinct disease phenotypes [1]. Studies on breast and other tumours suggested that classifiers based on a small set of marker genes can better classify tumour subtypes than can standard clinical criteria [2].

An obvious test statistic to employ in the detection of DE for a given gene is to form the usual two-sample (pooled) $t$-test statistic. However, as small class-sample sizes are often used for microarray experiments, there can be problems with a poor estimate of the variance in the denominator of the $t$-statistic if each gene is considered independently. One attempt 'at borrowing strength' across the tests is to adopt a moderated form of the $t$-statistic whereby the gene-specific variance is weighted with a contribution from all the genes; for example, [3]. Another commonly used attempt to involve all the gene profiles in the formation of a test statistic for a gene is to partition the gene profiles into clusters [4, 5, 6]. Existing clustering-based methods have been implemented by assuming the existence of pure clusters of up- and down-regulated genes, which are reflected by large class differences in the cluster-specific (estimated) mean expressions for the tissues [7, 8]. In an ideal situation, there would be three clusters corresponding to null genes, upregulated-DE genes, and down-regulaetd DE genes. But in practice, there is a need for more than three clusters since not all the up-regulated genes are assigned to the one cluster and, similarly, for the down-regulated genes. With more than three clusters there is the problem of how to identify the clusters corresponding null genes and up- and down-regulated DE genes, assuming the clusters are pure [9]. One way of approaching this problem has been to adopt an hierarchical approach whereby each cluster is decomposed into three subclusters representing null, up-regulated DE, and down-regulated genes [10].

Given pure clusters, the question of whether a gene is DE is decided on the basis of its (estimated) posterior probabilities of membership with respect to the various clusters. However, the intent of obtaining pure clusters is not always possible or verifiable. For example, a gene might have a high estimated posterior probability of belonging to a cluster taken to correspond to DE genes. But it can still be quite atypical of the cluster to which it is assigned and, indeed, be a null gene as to be illustrated below. The cluster-specific contrasts for tests on differences between the classes are the same for all genes in a cluster if no gene-specific differences are incldued in the contrast. They thus can be misleading.

In our approach, we therefore include gene-specific random effects terms in forming a cluster-specific contrast. Our initial clustering of the gene profiles into a number $g$ of clusters is effected by the fitting of a mixture of linear mixed models (LMMs) that include random effects terms specific to the genes in addition to the class-specific fixed effects terms. This model extends earlier work of [11] by allowing correlated gene-specific random effects for the $m$ tissue classes to follow a multivariate normal distribution, which gives a more flexible model with an additional correlation parameter explicitly governing and accounting for the relationship between gene expressions among the tissue classes due to individual gene effects; see Section 2. The component LMMs also include random effects terms shared by all genes belonging to the same component of the mixture model, which implies that these genes are not assumed to be independently distributed as usually assumed in other clustering or multiple hypothesis testing approaches; see also [12], [13], and [14]. The choice of the number $g$ of clusters is made by consideration of the likelihood via the BIC criterion [15]. However, the choice of $g$ is not crucial in that the subsequent use of them in our approach does not rely on the clusters being pure as to whether all cluster members are DE or not DE (null). In our approach, a (normalized) contrast for the test of no DE is formed based on class differences between the cluster-specific fixed effects terms for each class in addition to class differences between the gene-specific random effects terms. As discussed below, the inclusion of the latter terms in the cluster-specific contrast means that a gene not typical of the cluster to which has been assigned outright or in a soft manner by the initial clustering of the genes can modify the differences between the fixed effects terms for the cluster. The final form of our proposed test statistic is formed by weighting the cluster-specific (normalized) contrasts over the clusters. The test statistic can be used to rank the genes in order of evidence against the null hypothesis of no DE, which is often the primary aim of gene discovery experiemntns for which microarrays are designed. The distributional assumptions do not have to hold for the test statistic to peform well in ranking the genes. Our proposed test statistic can be used also to carry multiple hypothsis testing [16] where the intent is to control the false discovery rate (FDR) at or below a specified level; see also [17] and [18]. In the case of calculating the $P$-value, we adopt a permutation approach to provide an approximation to the null distribution of the test statistic and hence to the calculation of the associated $P$-value. The FDR can be controlled by using the [19] procedure or by, say, converting the $P$-vlaues for the genes to $z$-scores and then fitting mixtures of normal densities to these $z$-scores to obtain an estimate of the local false discovery rate in the form of the posterior probability that a gene is null given its $z$-score [20].

## 2 The model

We let $p_h$ be the number of samples in the $h$th tissue class $C_h$ $(h = 1, \ldots, m)$. The total number of measurements for each gene is $p = \sum_{h=1}^{m} p_h$. Thus the gene-expression data can be represented by a $n \times p$ matrix, where $n$ is the number of genes. We let $\boldsymbol{y}_j = (y_{1j}, \ldots, y_{pj})^T$ contain the measurements on the $j$th gene, where superscript $T$ denotes vector transpose; that is, $\boldsymbol{y}_j$ is the gene profile vector for the $j$th gene $(j = 1, \ldots, n)$. It is postulated that $\boldsymbol{y}_j$ has a $g$-component mixture distribution with probability $\pi_i$ of belonging to the $i$th component $G_i$ $(i = 1, \ldots, g)$, where the $\pi_i$ sum to one. We let the $g$-dimensional vector $\boldsymbol{z}_j$ denote the component membership of $\boldsymbol{y}_j$, where $z_{ij} = (\boldsymbol{z}_j)_i = 1$ if $\boldsymbol{y}_j$ belongs to the $i$th component and zero otherwise $(i = 1, \ldots, g)$. We put $\boldsymbol{y} = (\boldsymbol{y}_1^T, \ldots, \boldsymbol{y}_n^T)^T$ and $\boldsymbol{z} = (\boldsymbol{z}_1^T, \ldots, \boldsymbol{z}_n^T)^T$.

Conditional on its membership of the $i$th component $G_i$, the distribution of $\boldsymbol{y}_j$ is specified by the LMM [21]

$$\boldsymbol{y}_j = \boldsymbol{X}\boldsymbol{\beta}_i + \boldsymbol{U}\boldsymbol{b}_{ij} + \boldsymbol{V}\boldsymbol{c}_i + \boldsymbol{\epsilon}_{ij}, \tag{1}$$

where $\boldsymbol{X}, \boldsymbol{U}$, and $\boldsymbol{V}$ denote the design matrices corresponding, respectively, to the fixed effects terms $\boldsymbol{\beta}_i$ and to the random effects terms $\boldsymbol{b}_{ij}$ and $\boldsymbol{c}_i$ $(i = 1, \ldots, g)$.

The vector $\boldsymbol{b}_{ij} = (b_{1ij}, \ldots, b_{mij})^T$ contains the gene-specific random effects for each of the $m$ tissue classes, and $\boldsymbol{c}_i = (c_{1i}, \ldots, c_{pi})^T$ contains the random effects common to all genes from the $i$th component. The latter induces a correlation between those genes from the same component and is an attempt to allow for the fact that in reality the gene profile vectors are not all independently distributed. The gene-specific random effects $b_{hij}$ $(h = 1, \ldots, m)$ allow for correlation between the gene expressions across the tissues both within a class and between classes for the same gene. The expressions on a given gene should be independent but this may not hold in practice due to poor experimental conditions resulting in batch-effects.

The measurement error vector $\boldsymbol{\epsilon}_{ij}$ is taken to be multivariate normal $N_p(\boldsymbol{0}, \boldsymbol{A}_i)$, where $\boldsymbol{A}_i$ is a diagonal matrix. The vectors $\boldsymbol{b}_{ij}$ and $\boldsymbol{c}_i$ of random effects terms are taken to be multivariate normal $N_m(\boldsymbol{0}, \boldsymbol{B}_i)$ and $N_p(\boldsymbol{0}, \boldsymbol{C}_i)$, respectively, where $\boldsymbol{C}_i$ is assumed to be diagonal and $\boldsymbol{B}_i$ is a non-diagonal matrix given by

$$\boldsymbol{B}_i = \begin{pmatrix} \sigma_{b1i}^2 & \rho\sigma_{b1i}\sigma_{b2i} & \cdots & \rho\sigma_{b1i}\sigma_{bmi} \\ \rho\sigma_{b2i}\sigma_{b1i1} & \sigma_{b2i}^2 & \cdots & \rho\sigma_{b2i}\sigma_{bmi} \\ \vdots & \vdots & \vdots & \vdots \\ \rho\sigma_{bmi}\sigma_{b1i} & \rho\sigma_{bmi}\sigma_{b2i} & \cdots & \sigma_{bmi}^2 \end{pmatrix}, \tag{2}$$

where the additional parameter $\rho$ accounts for the correlation between gene-specific random effects $b_{hij}$ $(h = 1, \ldots, m)$, which are shared, respectively, among the expressions on the $j$th gene in the $h$th tissue class (see Supplementary materials for the covariance structure of the unconditional distribution of the gene-expression profiles under (1) and (2)).

## 2.1 The test statistic

We now define our test statistic for the ranking of the genes in order of their significance of being differentially expressed. We proceed initially under the assumption that each gene profile is classified with respect to the $g$ components $G_1, \ldots, G_g$ in its mixture distribution with components specified by (1); that is $z$ is known. We also assume initially that the vector $\zeta_i$ containing the distinct elements in the component-covariance matrices $A_i, B_i$, and $C_i$, is known.

We let $b_{G_i}$ be the $(mn_i)$-dimensional vector containing the random effects terms for the $n_i$ genes belonging to the $i$th component of the mixture model ($i = 1, \ldots, g$). We can write $b_{G_i}$ as

$$b_{G_i} = (b_{i_1}^T, \ldots, b_{i_{n_i}}^T)^T, \tag{3}$$

where $i_1, \ldots, i_{n_i}$ denote the labels of the $n_i$ genes belonging to the $i$th component $G_i$ ($i = 1, \ldots, g$) of the mixture model.

We let

$$r_i = (\beta_i^T, b_{G_i}^T, c_i^T)^T \quad (i = 1, \ldots, g) \tag{4}$$

be the vector containing the fixed and random effects for the $n_i$ genes belonging to the $i$th component $G_i$ ($i = 1, \ldots, g$). For an individual gene $j$ belonging to the $i$th component $G_i$, we can form the cluster-specific normalized contrast $S_{ij}$ given by

$$S_{ij} = d_j^T r_i / \lambda_{ij} \quad (i = 1, \ldots, g), \tag{5}$$

where $d_j$ is a vector whose elements sum to zero and $\lambda_{ij}$ is the normalizing term. The choice of $d_j$ has direct implication on the inference space of the contrast $S_{ij}$. For the case of $m=2$ classes of tissue samples, a typical form for $d_j$ is

$$d_j^T = (1 \ \text{-}1 \ \vdots \ 0 \ 0, \ \ldots, \ 0 \ 0, \ 1 \ \text{-}1, \ 0 \ 0, \ \ldots \ \vdots \ 0 \ \ldots \ 0), \tag{6}$$

where only one pair of (1 -1) exists in the second partition corresponding to the gene-specific random effects $b_{ij}$ for a gene in the $i$th cluster. The contrast (6) represents an "intermediate inference space" in that the inference is "narrow" to gene-specific random effects $b_{ij}$ but "broad" to tissue-specific random effects $c_i$ [22]. This means that a contrast of differential expressions between two classes of tissues is being considered and the inference applies to the specific genes studied in the experiment (narrow) and to the entire population from which biological tissue samples were obtained (broad).

Our proposed test statistic is based on an estimate $\hat{S}_{ij}$ of $S_{ij}$ obtained by replacing $r_i$ with an estimate $\hat{r}_i$ in the right-hand side of (5) and now taking the normalizing term $\lambda_{ij}$ to be the standard error of $d_j^T \hat{r}_i$ (conditional on membership of the $j$th gene to the $i$th component $G_i$). That is,

$$\hat{S}_{ij} = d_j^T \hat{r}_i / \lambda_{ij}, \tag{7}$$

where $\hat{r}_i$ is an estimate of the vector $r_i$ of fixed and random effects terms and

$$\lambda_{ij} = \sqrt{d_j^T \operatorname{cov}(\hat{r}_i) d_j}. \tag{8}$$

In the case where $\hat{\boldsymbol{r}}_i$ is the BLUP estimator of $\boldsymbol{r}_i$,

$$\text{cov}(\hat{\boldsymbol{r}}_i) = \boldsymbol{\Omega}_i(\boldsymbol{\zeta}_i; \boldsymbol{z}), \tag{9}$$

where $\boldsymbol{\Omega}_i$ is defined as follows.

The matrix $\boldsymbol{\Omega}_i$ (and $\boldsymbol{d}_j$) is partitioned conformally corresponding to $\boldsymbol{\beta}_i|\boldsymbol{b}_{G_i}|\boldsymbol{c}_i$ with dimensions $m$, $mn_i$, and $p$, respectively. That is,

$$\boldsymbol{\Omega}_i = \begin{bmatrix} \boldsymbol{\Omega}_{i\beta} & \boldsymbol{\Omega}_{i\beta b} & \boldsymbol{\Omega}_{i\beta c} \\ \boldsymbol{\Omega}_{i\beta b}^T & \boldsymbol{\Omega}_{ib} & \boldsymbol{\Omega}_{ibc} \\ \boldsymbol{\Omega}_{i\beta c}^T & \boldsymbol{\Omega}_{ibc}^T & \boldsymbol{\Omega}_{ic} \end{bmatrix}^{-1}, \tag{10}$$

where $\boldsymbol{\Omega}_{i\beta} = n_i \boldsymbol{X}^T \boldsymbol{A}_i^{-1} \boldsymbol{X}$, $\boldsymbol{\Omega}_{i\beta b} = \boldsymbol{1}_{n_i} \otimes \boldsymbol{X}^T \boldsymbol{A}_i^{-1} \boldsymbol{U}$,

$$\boldsymbol{\Omega}_{i\beta c} = n_i \boldsymbol{X}^T \boldsymbol{A}_i^{-1} \boldsymbol{V}, \quad \boldsymbol{\Omega}_{ib} = \boldsymbol{I}_{n_i} \otimes (\boldsymbol{U}^T \boldsymbol{A}_i^{-1} \boldsymbol{U} + \boldsymbol{B}_i^{-1}),$$

$$\boldsymbol{\Omega}_{ibc} = \boldsymbol{1}_{n_i} \otimes \boldsymbol{U}^T \boldsymbol{A}_i^{-1} \boldsymbol{V}, \quad \boldsymbol{\Omega}_{ic} = n_i (\boldsymbol{V}^T \boldsymbol{A}_i^{-1} \boldsymbol{V} + \boldsymbol{C}_i^{-1}), \tag{11}$$

where $\boldsymbol{1}_{n_i}$ is a $n_i$-dimensional vector of ones, $\boldsymbol{I}_{n_i}$ is an identity matrix with dimension $n_i$, and the sign $\otimes$ denotes the Kronecker product of two matrices.

The computation of the inverse matrix (10) is not straightforward as the partition corresponding to $\boldsymbol{b}_{G_i}$ involves a large dimensional block matrix (see the Supplementary materials for the derivation of the inverse matrix (10) in terms of (11)). In evaluating $\hat{\boldsymbol{\Omega}}_i$ in (10), it is noted that the empirical approximation, using the maximum likelihood (ML) estimates $(\hat{\boldsymbol{A}}_i, \hat{\boldsymbol{B}}_i, \hat{\boldsymbol{C}}_i)$, tends to underestimate the true sampling variability of the estimated contrast $\hat{S}_{ij}$ of mixed effects. It is because the uncertainty in estimating the variance components is not accounted for. Based on a nonparametric bootstrap method [23], we found that the variances of the variance component estimates are generally small (contributing less than 5% of variance component estimates). Thus, the use of estimated variance components does not appear to introduce a large bias on $\hat{\boldsymbol{\Omega}}_i$; see [22].

Now in practice we do not know the classification $\boldsymbol{z}$ of the gene profiles with respect to the $g$ components $G_1, \ldots, G_g$ nor the vector $\boldsymbol{\zeta}_i$ of variances/covariances. We therefore fit a mixture of $g$-component LMM distributions as defined by (1) with the value of $g$ is chosen according to the BIC criterion. The vector $\boldsymbol{\Psi}$ containing the unknown mixing proportions $\pi_i$, the fixed effects $\boldsymbol{\beta}_i$, and the vector $\boldsymbol{\zeta}_i$ of variances/covariances ($i = 1, \ldots, g$) is estimated by ML. The random effects terms $\boldsymbol{b}_{G_i}$ and $\boldsymbol{c}_i$ are estimated by $\hat{\boldsymbol{b}}_{G_i} = E_{\widehat{\boldsymbol{\Psi}}}\{\boldsymbol{b}_{G_i} \mid \boldsymbol{y}_j, z_{ij} = 1\}$ and $\hat{\boldsymbol{c}}_i = E_{\widehat{\boldsymbol{\Psi}}}\{\boldsymbol{c}_i \mid \boldsymbol{y}\}$, respectively, where $E_{\widehat{\boldsymbol{\Psi}}}$ denotes expectation using $\widehat{\boldsymbol{\Psi}}$ for $\boldsymbol{\Psi}$. For unknown $\boldsymbol{\zeta}_i$ and $\boldsymbol{z}$, we make the approximation

$$\text{cov}(\hat{\boldsymbol{r}}_i) \approx \boldsymbol{\Omega}_i(\widehat{\boldsymbol{\zeta}}_i, \hat{\boldsymbol{z}}); \tag{12}$$

see the Supplementary materials for the justification of this approximation.

Conditional on membership of the $j$th gene from component $G_i$, this leads to the estimated (normalized) contrast,

$$\hat{S}_{ij} = \boldsymbol{d}_j^T \hat{\boldsymbol{r}}_i / \sqrt{\boldsymbol{d}_j^T \boldsymbol{\Omega}_i(\widehat{\boldsymbol{\zeta}}_i; \hat{\boldsymbol{z}}) \boldsymbol{d}_j}, \tag{13}$$

where
$$\hat{\boldsymbol{r}}_i = (\hat{\boldsymbol{\beta}}_i^T, \hat{\boldsymbol{b}}_{G_i}^T, \hat{\boldsymbol{c}}_i^T)^T. \tag{14}$$

On weighting now the estimated (normalized) contrast $\hat{S}_{ij}$ over the $g$ components in the mixture model, we obtain as our test statistic for the $j$th gene,

$$W_j = \sum_{i=1}^{g} \tau_i(\boldsymbol{y}_j; \widehat{\boldsymbol{\Psi}}, \hat{\boldsymbol{c}}_i) \hat{S}_{ij}, \tag{15}$$

where $\tau_i(\boldsymbol{y}_j; \boldsymbol{\Psi}, \boldsymbol{c}_i)$ is the posterior probability that the $j$th gene belongs to the $i$th component $G_i$ conditional on $\boldsymbol{y}_j$ and $\boldsymbol{c}_i$.

## 2.2 Null Distribution of $W_j$

The weighted contrast $W_j$ can be formed to test the null hypothesis of $H_j$ : $j$th gene is not DE ($j = 1, \ldots, n$). The null distribution of $W_j$ can be assessed using a permutation method described as follows. The steps are:

1. Let $\boldsymbol{H}$ be the $n \times p$ data matrix, where $\boldsymbol{H} = (\boldsymbol{y}_1 \cdots \boldsymbol{y}_n)^T$. Permute the class labels $B$ times (corresponding to the $p$ columns of $\boldsymbol{H}$) and let $\boldsymbol{y}_j^{(b)}$ be the gene profile for the $j$th gene after the $b$th permutation ($b = 1, \ldots, B; j = 1, \ldots, n$). Then after the $b$th permutation we have corresponding to $\boldsymbol{H}$, the data matrix

$$\boldsymbol{H}^{(b)} = (\boldsymbol{y}_1^{(b)} \cdots \boldsymbol{y}_n^{(b)})^T.$$

2. Then for each gene $j$ in turn, compute the $B$ replications $W_j^{(b)}$ of $W_j$ ($b = 1, \ldots, B$), where $W_j^{(b)}$ is calculated in the same way that $W_j$ is calculated except that $\boldsymbol{y}_j$ is replaced by $\boldsymbol{y}_j^{(b)}$. That is, at the end of the estimation process for the original data $\boldsymbol{H}$, we compute the $W_j$. Now to compute the $W_j^{(b)}$ we repeat the last process but with $\boldsymbol{H}$ replaced by $\boldsymbol{H}^{(b)}$.

As stated above, it follows from (13) to (15) that $W_j^{(b)}$ is given by

$$W_j^{(b)} = \sum_{i=1}^{g} \tau_i(\boldsymbol{y}_j^{(b)}; \widehat{\boldsymbol{\Psi}}, \hat{\boldsymbol{c}}_i) \hat{S}_{ij}^{(b)}, \tag{16}$$

where
$$\hat{S}_{ij}^{(b)} = \boldsymbol{d}_j^T \hat{\boldsymbol{r}}_i^{(b)} / \sqrt{\boldsymbol{d}_j^T \boldsymbol{\Omega}_i(\widehat{\boldsymbol{\zeta}}_i; \hat{\boldsymbol{z}}) \boldsymbol{d}_j} \tag{17}$$

and where
$$\hat{\boldsymbol{r}}_i^{(b)} = (\hat{\boldsymbol{\beta}}_i^T, \hat{\boldsymbol{b}}_{G_i}^{(b)T}, \hat{\boldsymbol{c}}_i^T)^T. \tag{18}$$

In (18), we have
$$\hat{\boldsymbol{b}}_{G_i}^{(b)} = E_{\widehat{\boldsymbol{\Psi}}}\{\boldsymbol{b}_{G_i} \mid \boldsymbol{y}_j^{(b)}, z_{ij} = 1\}. \tag{19}$$

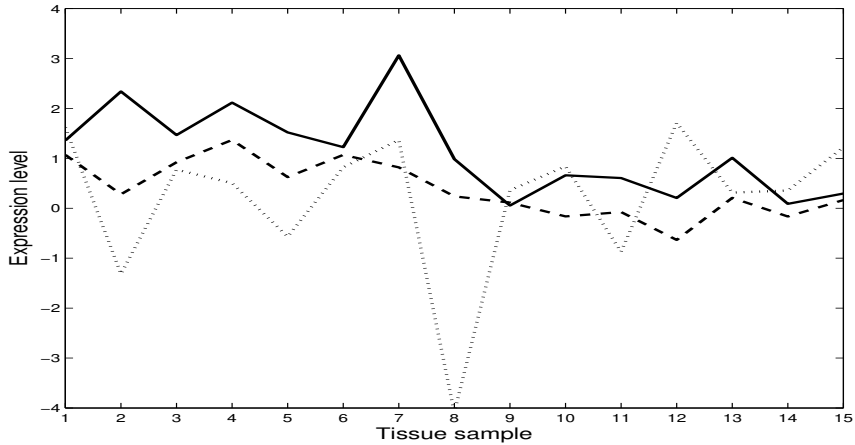

Figure 1: Three selected gene profiles (7 BRCA1- and 8 BRCA2-mutation-positive tumour tissue samples): Solid line (from Cluster 3, $\hat{\boldsymbol{\beta}}_{13} - \hat{\boldsymbol{\beta}}_{23} = 0.303$, difference in gene-specific random effects $\hat{b}_{13j} - \hat{b}_{23j} = 0.646$); Dashed line (from Cluster 4, $\hat{\boldsymbol{\beta}}_{14} - \hat{\boldsymbol{\beta}}_{24} = 0.047$, $\hat{b}_{14j} - \hat{b}_{24j} = 0.487$); Dotted line (from Cluster 1, $\hat{\boldsymbol{\beta}}_{11} - \hat{\boldsymbol{\beta}}_{21} = -0.351$, $\hat{b}_{11j} - \hat{b}_{21j} = 0.387$).

## 2.3  Data Preprocessing

In practice, it is common to standardize each column of the data matrix to have zero mean and unit standard deviation, followed by the row standardization to shift expression profiles to the same baseline (zero mean, unit standard deviation) for comparison. However, row standardization will transform the data into directional data that lie on a unit hypersphere [24]. The latter implies that multivariate normal distributions may not be appropriate for modelling row-standardized gene-expression profiles of high dimensions [25]. In this paper, data preprocessing is implemented by column standardization only, which helps to standardize the variability of gene expressions for each tissue sample and hence facilitates the clustering of genes. Without row standardization, it implies that genes with different mean expressions shall be clustered into different clusters, despite the fact that they may exhibit similar differentiation across the tissue classes. In other words, genes within the same cluster shall have different patterns of differential expression. A cluster of genes may therefore contain both DE and non-DE genes; see Section 3. This is consistent with the approach presented in [10] where three patterns of differential expression are nested within a mixture model.

## 3  Results

Using three published gene-expression data sets, we first show how the proposed method can be used to detect DE genes between two tissue classes. Results are pre-sented for functional annotation and enrichment analysis of those top-ranked genes. We then compare our method with different approaches using a benchmark data set in terms of the false discovery proportion [26] for the top ranked genes. Additional empirical studies using simulated data are also performed to illustrate advantages of our method compared to existing methods applied in a multiple hypothesis testing context with the FDR controlled at a specified level.

### 3.1 Identification of differentially-expressed genes

We applied the clustering-based contrast method to the breast cancer dataset of [27]. The data comprised the measurement of 3226 genes using cDNA arrays, for $p_1=7$ BRCA1-mutation-positive tumours and $p_2=8$ BRCA2-mutation-positive tumours. The aim is to identify DE genes between the tumours associated with the different mutations. We column normalized the logged expressions and fitted the proposed extended random-effects model to the data with $g=3$ to $g=15$ clusters. Using BIC, we identified five clusters of genes (see Supplementary materials for details of the clustering results including the ML estimates of the unknown parameters). The difference $\hat{\boldsymbol{\beta}}_{1i} - \hat{\boldsymbol{\beta}}_{2i}$ between the estimated mean expressions in Class 1 and Class 2 for the $i$th component are -0.351, -0.131, 0.303, 0.047, and -0.147 for $i = 1, \ldots, 5$, respectively. However, it is not necessary all genes in the clusters with large absolute differences $\hat{\boldsymbol{\beta}}_{1i} - \hat{\boldsymbol{\beta}}_{2i}$ (such as for $i = 1$ or 3) are DE. On the other hand, a DE gene $j$ may be grouped into Cluster 4, but can have a large weighted contrast due to a large difference $\mid \hat{b}_{1ij} - \hat{b}_{2ij} \mid$ between the (estimated) gene-specific random effects terms for $i=4$. As an illustration, we depict in Figure 1 three gene profiles. Two of them are among the top 100 ranked genes, but one is from Cluster 3 (large absolute difference $\mid \hat{\boldsymbol{\beta}}_{13} - \hat{\boldsymbol{\beta}}_{23} \mid$) and the other is from Cluster 4 (small absolute difference $\mid \hat{\boldsymbol{\beta}}_{14} - \hat{\boldsymbol{\beta}}_{24} \mid$). The remaining gene profile may correspond to a non-DE gene which, however, would be assigned to Cluster 1 with large absolute difference $\mid \hat{\boldsymbol{\beta}}_{11} - \hat{\boldsymbol{\beta}}_{21} \mid$) (see also Supplementary materials, Table S2).

The ranking of DE genes is implemented on the basis of the weighted estimates of contrast of mixed effects (15). The top 50 ranked up-regulated genes (corresponding to large negative weighted contrast) and the top 50 down-regulated genes (corresponding to large positive weighted contrast) in BRCA2-mutation-positive tumours relative to tumours with BRCA1 mutations are presented for functional annotation and enrichment analyses, using web-based DAVID Functional Annotation Tool [28] and GeneGo MetaCore pathway analysis program.

### 3.2 Functional annotation and enrichment analyses

We compare our clustering-based contrast method with a multiple hypothesis testing approach that is based on the pooled two-sample $t$-statistic [2, 29]. The gene lists obtained by both methods (contrast and $t$-test methods) were submitted to DAVID Bioinformatics Resources 6.7 for Functional Annotation analyses. Gene Ontology (GO) terms with the most significant gene enrichment statistics were selected for each method and compared. The same gene lists with mean expression values were also evaluated by "compare experiment" module of the GeneGo MetaCore pathway analysis tool (Thomson Reuters, St. Joseph, MI) for reports of common and unique networks identified by the two methods.

The results of the functional annotation and enrichment analyses by DAVID Functional Annotation Tool are presented in Table 1. It can be seen that our top-ranked genes are involved in more annotated functions with slightly higher significance. In the analysis by [27], coordinated transcription activation of genes involved in DNA repair and induction of apoptosis was found in tumours with BRCA1 mutations, and genes involved in suppression of apoptosis in tumours with BRCA2 mutations (as reference). Our clustering-based contrast method has identified these characteristics in BRCA1- and BRCA2-mutation-positive tumours, respectively. In addition, our method has also identified up-regulated genes involved in new significant pathways represented by GO terms such as "blood vessel morphogenesis" and "response to wounding" in tumours with BRCA2 mutations, as well as "regulation of mRNA stability" and "cellular protein metabolic processes" in tumours with BRCA1 mutations. We has also used GeneGo MetaCore pathway analysis tool to compare network enrichment of genes identified by the two methods (see Supplementary materials, Table S3). The genes identified by the contrast method show significant enrichment in genes that regulate cell cycle and cell adhesion functions. These new findings may provide further insights into the differences between BRCA1- and BRCA2-mutated tumours.

Another illustration using the breast cancer dataset of [30] is given in the Supplementary materials.

### 3.3 Golden Spike benchmark dataset

Golden Spike is a dataset generated to provide a benchmark for comparing different approaches for the analysis of Affymetrix GeneChips [31]. The experiment compared two classes of tissue samples: $p_1 = 3$ (control) and $p_2 = 3$ (spike-in). By design, the data set has 1331 (9.5%) DE (nonnull) genes (nominal spike-in to control ratio > 1) and 12666 (90.5%) true nulls (nominal spike-in to control ratio = 1 or do not matched to any cRNA). In this study, the log-transformed data set "10a" is used, which is available in the website of [31]; see also [32]. We fitted the extended random-effects model to the data and identified four clusters of genes based on BIC for model selection. Figure 2 presents the proportion of true nulls among the top-ranked genes at various cut-off rankings. We compare our method with existing approaches including Significance Analysis of Microarray (SAM) [29], Linear Models for Microarray Data (LIMMA) [3], Optimal Discovery Procedure (ODP) [33], and the $t$-test. From Figure 2, it can be seen that the proposed method outperforms the other approaches by having the smallest proportion of true nulls when the cut-off ranking is larger than 700. When the cut-off ranking is small, the proposed method performs equally well as ODP. Among the top 1000 ranked genes (declared to be DE),

Table 1: Comparison of GO term association of the top 50 ranked up- and down-regulated genes in tumours with BRCA2 mutations relative to tumours with BRCA1 mutations (breast cancer data of [27])

|  | contrast | | $t$-test | |
|---|---|---|---|---|
| GO terms associated with up-regulated genes | ct (fold) | $P$-value | ct (fold) | $P$-value |
| Response to hormone stimulus | 9 (7.9) | 1.2E-05 | 6 (5.5) | 3.9E-03 |
| Cell motion | 8 (5.4) | 5.0E-04 | 3 (2.1) | 4.0E-01 |
| Enzyme linked receptor protein signal pathway | 7 (6.6) | 5.3E-04 | 6 (5.9) | 2.9E-03 |
| Regulation of cell proliferation | 9 (3.7) | 2.2E-03 | 8 (3.4) | 6.6E-03 |
| Regulation of phosphorylation | 7 (4.8) | 2.6E-03 | 5 (3.6) | 4.4E-02 |
| Transcription factor binding | 7 (4.5) | 3.5E-03 | 4 (2.7) | 1.7E-01 |
| Negative regulation of cell differentiation | 5 (7.5) | 4.0E-03 | 4 (6.3) | 2.4E-02 |
| Cell-matrix adhesion | 3 (11) | 3.0E-02 | 4 (15) | 2.1E-03 |
| Response to drug | 7 (10) | 4.4E-05 | | |
| Negative regulation of apoptosis | 7 (6.4) | 6.4E-04 | | |
| Blood vessel morphogenesis | 5 (7.6) | 3.7E-03 | | |
| Response to wounding | 7 (4.3) | 4.9E-03 | | |
| Cellular response to stress | | | 6 (3.7) | 2.2E-02 |
| | | | | |
| GO terms associated with down-regulated genes | | | | |
| Interphase of mitotic cell cycle | 6 (18) | 1.6E-05 | 6 (17) | 2.2E-05 |
| Cyclin-dependent protein kinase activity | 5 (29) | 2.3E-05 | 4 (22) | 7.6E-04 |
| Nuclear lumen | 13 (4) | 1.6E-04 | 13 (3) | 7.0E-04 |
| Posttranscriptional regulation - gene expression | 6 (9.0) | 4.7E-04 | 5 (7.0) | 5.2E-03 |
| Regulation to DNA damage stimulus | 6 (5.1) | 5.7E-03 | 9 (7.1) | 2.7E-05 |
| Induction of apoptosis | 3 (3.0) | 2.6E-01 | 5 (4.6) | 2.1E-02 |
| Regulation of mRNA stability | 3 (43) | 2.1E-03 | | |
| Cellular protein metabolic process | 5 (8.7) | 2.3E-03 | | |
| Single-stranded DNA binding | | | 5 (28) | 2.6E-05 |

Note: ct is the number of genes belonging to an annotation term, fold is the fold-enrichment of involved genes over total genes in the list relative to the number of genes with the same term category in the human genome background, $P$-value is the modified Fisher exact $P$-value for assessing the gene-enrichment; see [28]

our method gives the least of 79 true nulls. The numbers of true nulls for the other approaches are, respectively, 311 (SAM), 141 (SAM with its fudge factor $s0$ set to zero), 290 (LIMMA), 227 ($t$-test), and 140 (ODP).

## 3.4 Simulation experiment 1: ranking of correlated genes

We consider the simulation procedure described in [34] to investigate the relative performance of the contrast method when gene expressions are correlated. For $m=2$ classes of tissues with $p_1=p_2=10$, we generated 3000 gene expression levels indepen-dently for each of the 20 tissues in six blocks of size 500 from multivariate normal distributions with mean vector $10 \times \mathbf{1}_{500}$ and covariance matrix $\sigma^2\{\mathbf{1}_{500}\mathbf{1}_{500}^T\rho + (1-\rho)\mathbf{I}_{500}\} \otimes \mathbf{I}_6$. Here, $\sigma^2=4$ is the common variance and $\rho$ is the correlation between gene expression levels. The above covariance structure implies that genes in the same block are correlated, while genes in different blocks are independent. Finally, for 20% of the genes (300 randomly selected genes each for up- and down-regulated gene groups), a true mean difference in expression $\pm\delta$ between the two tissue classes was added to the expression levels for the last 10 tissues.

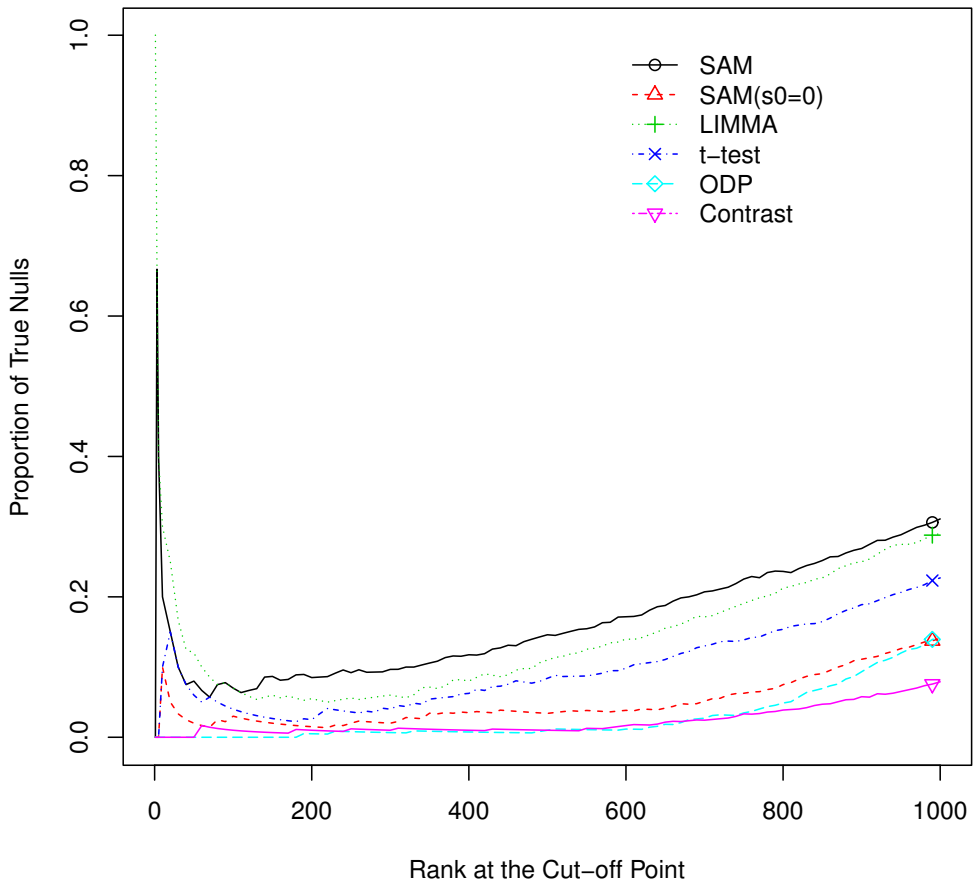

Figure 2: Proportion of true nulls among the top-ranked genes.

We consider $\rho=0.0$ (independence), $\rho=0.4$ (moderate), $\rho=0.6$ (moderately strong), $\rho=0.8$ (strong dependence), and $\delta=2$ (moderate differential expression). For each set of parameter values, 100 independent simulation experiments were conducted. We fitted the extended random-effects model to the column-normalized data with $g = 3$ components. Given that there are 600 DE genes from a total of $n=3000$ genes in each simulated data set, we obtained for each method the top ranked 600 genes and noted the proportion of null genes among them; that is, the false discovery proportion (FDP), and the proportion of the 600 DE genes among them (the power). Figure 3 presents the results comparing the contrast method with the $t$-test, SAM, LIMMA, and ODP. The proposed contrast method again outperforms the other approaches with higher power and a markedly smaller error rate in FDP. It should be noted that ODP was shown by [33] to outperform most of the leading methods for identifying DE genes [35].

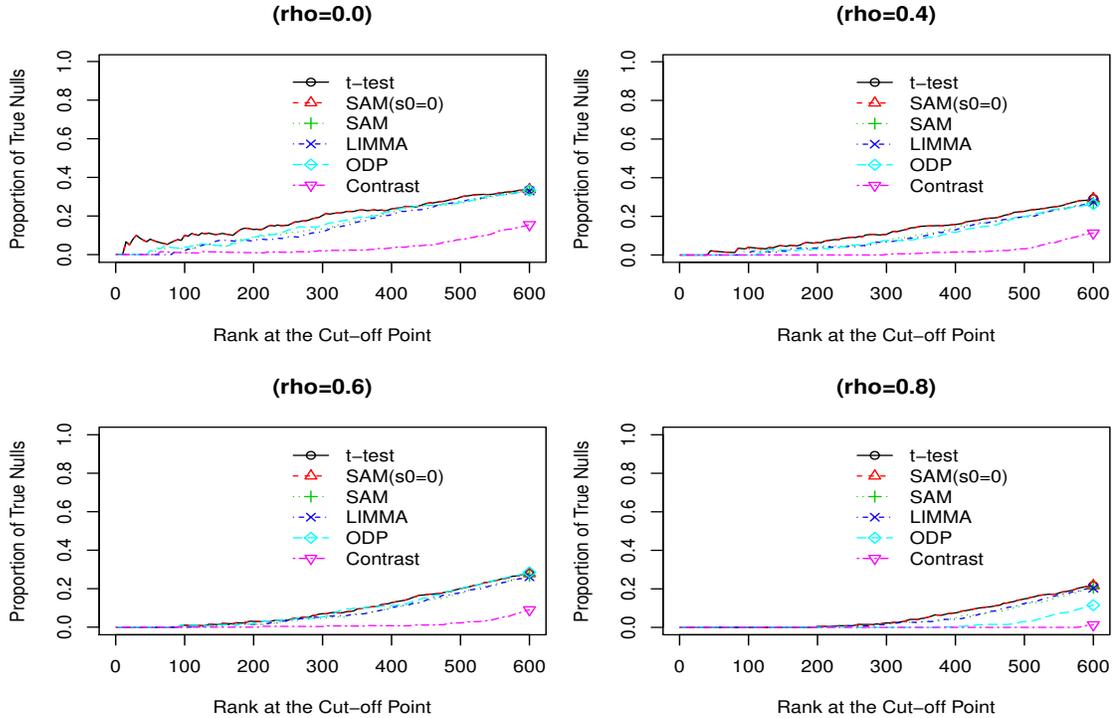

Figure 3: Comparative study on a synthetic microarray data with correlated gene expressions: true FDP among the top-ranked genes for Set 1: $\rho$=0.0, $\delta$=2; Set 2: $\rho$=0.4, $\delta$=2; Set 3: $\rho$=0.6, $\delta$=2; and Set 4: $\rho$=0.8, $\delta$=2.

### 3.5 Simulation experiment 2: controlling the FDP

In this simulation experiment, we consider $m$=2 two classes of tissues and the same setting as in Simulation experiment 1 above. Four sets of parameter values of $(\delta, \rho)$ are considered: ($\delta$=2, $\rho$=0.0), ($\delta$=2, $\rho$=0.4), ($\delta$=3, $\rho$=0.0), and ($\delta$=3, $\rho$=0.4).

In this experiment we illustrate our contrast method, where the intent is to control the FDR. One way to achieve this is to use the [19] procedure applied to the $P$-values. In order to be able to compute a $P$-value for the test statistic $W_j$ of our contrast methods, we need to know its null distribution under the null hypothesis that the $j$th gene is not DE. A rough approximation would be to take $W_j$ as having a standard normal distribution. We use an improved approximation by adopting a $t_v$ distribution with degrees of freedom $\nu$ estimated by fitting this $t$-distribution to some replicated null values of $W_j$ obtained by permuting the column labels as detailed in Section 2.2. Having obtained $P$-values for each gene, we can proceed as in [20] and convert them to $z$-scores to which a mixture of two normals is fitted with the first component corresponding to the null genes and the second component corresponding to the DE genes. The (estimated) posterior probability $\hat{\boldsymbol{\tau}}_0(z_j)$ that the $j$th genes belongs to the first (null) component is the estimate of the local FDR. With each method, genes having an estimated local FDR less than some threshold $c_0$ (that is, $\hat{\boldsymbol{\tau}}_0(z_j) < c_0$) can be taken to be DE. The implied FDR and power can be estimated as outlined in Section 2. In Table 2, the true FDP and power are presented. It can be seen that the contrast method has high power when there is moderate differential expression or moderate dependence. The other methods have less power (see Simulation experiment 1) and their results are not included in Table 2.

Table 2: Simulation study on controlling the FDR (contrast method)

| Set | $N_r$ | True FDP | True FNDP | True power |
|---|---|---|---|---|
| 1 ($\delta=2$, $\rho=0$) | 502 | 0.0817 | 0.0556 | 0.7683 |
| 2 ($\delta=2$, $\rho=0.4$) | 541 | 0.0610 | 0.0374 | 0.8467 |
| 3 ($\delta=3$, $\rho=0$) | 601 | 0.0266 | 0.0062 | 0.9750 |
| 4 ($\delta=3$, $\rho=0.4$) | 609 | 0.0214 | 0.0017 | 0.9933 |

Note: The value $N_r$ is the number of selected genes for differential expression

# 4 Discussion and Conclusion

We have presented a clustering-based contrast approach to draw inference on differences between classes using full gene-expression profiles. An extended random-effects model is adopted for the clustering of gene-expression data, which allows for a direct modelling of correlation among genes and within genes via correlated gene-specific random effects. Our approach thus enables the partition of overall variation into correlated random components and independent random noise. More importantly, the predicted random effects have a meaningful interpretation and, together with the fixed effects, they can be adopted to form a weighted contrast for assessing directly the differential expression between tissue classes for each gene. The applicability of the proposed method has been demonstrated using three published real data sets. The results show that there is significant enrichment of highly-ranked genes that are associated with particular biological functions such as cell division and cell proliferation. Empirical comparisons presented in the Results section demonstrate that the proposed inference method, with the use of full gene-expression profiles, has higher power to detect DE genes and outperforms multiple hypothesis testing approaches.

In applications where the attempt is to identify marker genes for accurate classification of disease subtypes [30], simply picking the top-ranked genes may not be efficient as the selected genes could be highly correlated among themselves [36]. Including lots of these redundant genes may confuse classifiers, with the consequent possibility of misleading classifications being made [2]. The proposed clustering-based contrast approach opens up a new way within a minimum redundancy and maximum relevance framework [36] to form a list of marker genes useful for better classifications of disease phenotypes. As correlated genes are grouped into the same cluster, a few top-ranked genes selected from each cluster may provide a more complementary list of marker genes that capture broader characteristics of disease phenotypes. The relative performance of this approach in constructing a classifier for the prediction of disease subtypes will be pursued in future research.

# Acknowledgement

This work was supported by the Australian Research Council.